\documentstyle[12pt]{article}
\newlength{\dinwidth}
\newlength{\dinmargin}
\newcommand{\ba}{\begin{array}}
\newcommand{\ea}{\end{array}}
\newcommand{\be}{\begin{equation}}
\newcommand{\ee}{\end{equation}}
\newcommand{\bea}{\begin{eqnarray}}
\newcommand{\eea}{\end{eqnarray}}

\def\bee{\begin{eqnarray}}
\def\eee{\end{eqnarray}}
\def\be{\begin{equation}}
\def\ee{\end{equation}}

\begin{document}
\thispagestyle{empty}
\addtocounter{page}{-1}
\begin{flushright}
IASSNS HEP 98-84\\
SNUTP 98 / 121\\
{\tt hep-th/9812062}\\
\end{flushright}
\vspace*{1.3cm}
\centerline{\Large \bf Green-Schwarz Superstring on $AdS_3 \times S^3
$~\footnote{
Work supported in part by the U.S. Department of Energy under Grant No.
DE-FG02-90-ER40542, KOSEF Interdisciplinary Research Grant and SRC-Program,
KRF International Collaboration Grant, Ministry of Education Grant BSRI 
98-2418, SNU Faculty Research Grant, and The Korea Foundation for Advanced 
Studies Faculty Fellowship.}}
\vspace*{1.2cm} \centerline{\large \bf
Jaemo Park${}^a$ and Soo-Jong Rey${}^b$}
\vspace*{0.8cm}
\centerline{\large\it School of Natural Sciences, Institute for Advanced
Study}

\vskip0.3cm
\centerline{\large\it Olden Lane, Princeton NJ 08540 USA${}^a$}
\vspace{0.5cm}
\centerline{\large\it Physics Department \& Center for Theoretical Physics}
\vskip0.3cm
\centerline{\large\it Seoul National University, Seoul 151-742 KOREA${}^b$}
\vskip0.8cm
\centerline{\tt jaemo@sns.ias.edu, \quad sjrey@gravity.snu.ac.kr}
\vspace*{1.5cm}
\centerline{\Large\bf abstract}
\vspace*{0.5cm}
Green-Schwarz action of Type IIB string on $AdS_3 \times S^3$ is constructed
via coset superspace approach. The construction relies exclusively on
symplectically Majorana-Weyl spinor formalism, thus permitting it easier 
to prove $\kappa$-symmetry for on-shell Type IIB supergravity backgrounds.
\vspace*{1.1cm}

\baselineskip=20pt
\newpage

\section{Introduction}
\setcounter{equation}{0}
Recently, prompted by the observation that anti-de Sitter supergravity is
dual to conformal field theory at large $N$ and large `t Hooft coupling 
\cite{malda}, there has arisen renewed interest to study string dynamics in 
curved spacetime. According to the correspondence, string worldsheet and 
loop corrections correspond to inverse `t Hooft coupling and $1/N$ 
corrections in the dual conformal field theory. As such, it should be important 
to develop a systematic method that would enable construction of superstring 
action on a curved background and quantization thereof. For the anti-de 
Sitter spaces under consideration, the background curvature is induced 
by Ramond-Ramond tensor field flux threaded through compact space. This has 
led one to investigate string and M-theory on anti-de Sitter space, 
including fundamental string \cite{MeTs,kallosh} or $p$-brane propagation 
\cite{adsdbrane, plefka} and Type IIB $SL(2,{\bf Z})$ duality \cite{duality},
a subject which has been left largely unexplored until recently.

For the background $AdS_5 \times S^5$, a Green-Schwarz superstring action has 
been constructed recently\cite{MeTs}. The construction is based on supercoset 
space approach and exhibits manifest $\kappa$-symmetry for on-shell Type 
IIB supergravity backgrounds. Although this marks a significant progress, 
$\kappa$-symmetry gauge-fixing and quantization thereof still seem to pose 
considerable difficulties. Therefore, it would be most desirable to study the 
superstring propagation on a simpler Ramond-Ramond background.

One of the simpler backgrounds is Type IIB string on $AdS_3 \times S^3
\times M_4$, where $M_4$ denote a Ricci-flat four-dimensional compact manifold
such as $T^4$ or $K3$. In fact, this background arises as the near-horizon
geometry of Type IIB D1-D5 brane configuration \cite{malda} (wrapped on $T^4$ 
or $K_3$), which threads nonzero Ramond-Ramond (RR) tensor flux through $M_4$. 
By performing $SL(2,{\bf Z})$ S-duality rotation, which is now part of 
U-duality group in six-dimensions, the configuration is mapped  
to Type IIB F1-NS5 brane configuration \cite{giveonkutasovseiberg}. The 
six-dimensional background is now $AdS_3 \times S^3$ threaded by (Neveu-Schwarz)-(Neveu-Schwarz) (NS-NS) tensor field flux. 
For $M^4 = T^4$ or $K^3$, on which the D1-D5 or F1-NS5 branes are 
wrapped on, the six-dimensional superstring carries sixteen supersymmetries, 
half as much as the ten-dimensional Type IIB superstring. Clearly, from 
six-dimensional $(2,0)$ supergravity point of view, both cases correspond to 
a consistent on-shell background, on which a non-critical superstring can 
propagates. One novelty of these backgrounds is that the symmetry 
group $SO(2,2)$ of the $AdS_3$ space is realized on the dual conformal field 
theory as an infinite-dimensional Virasoro algebra. 

Recently, the Green-Schwarz superstring action on $AdS_3 \times S^3$ has been 
constructed by Pesando and by Rahmfeld and Rajaraman \cite{pesando}. 
It turns out that their formulations are 
not the most convenient ones for investigation of certain interesting issues
such as consistency check with on-shell background of six-dimensional $(2,0)$ 
supergravity \cite{romans}, which has been formulated exclusively using the 
symplectically Majorana-Weyl $d=2$spinors, and U-duality transformation.  

In the present paper, utilizing symplectically Majorana-Weyl $d=6$ spinor 
formalism, we provide a new formulation of the Green-Schwarz superstring 
action on $AdS_3 \times S^3$. As shown in section 2, the symplectically 
Majorana-Weyl $d=6$ spinors descends in fact from the Majorana-Weyl $d=10$ 
spinors when decomposed on the above background. With the new formalism, one 
can prove in a straightforward way the $\kappa$-symmetry constraints to 
on-shell $(2,0)$ supergravity background.
The formalism also enables to investigate straightforwardly the 
Type IIB $SL(2,{\bf Z})$ duality transformation, which is now a subgroup
of six-dimensional U-duality transformation.
In section 3, employing the coset superspace approach, we then construct
the Green-Schwarz action for a fundamental (non-critical) string 
propagating on $AdS_3 \times S^3$, on which nonzero R-R (but not NS-NS ) 
tensor field flux is threaded.

\section{Symplectically Majorana-Weyl Spinor Decomposition}
Before we dwell on construction of the Green-Schwarz superstring action, it 
is useful to recapitulate decomposition of the $d=10$ Type IIB spinors into 
spinors on the product space under consideration~\footnote{Related analysis 
have been considered in \cite{boonstra, kallosh3}.}. In the Type IIB string, 
there are thirty-two supercharges, consisting of two sixteen component, 
Majorana-Weyl spinors $\eta_L, \eta_R$. Here, $L,R$ subscript denotes their 
origin from the left- and the right-movers on the world-sheet, viz., 
the spacetime spinors are counted after chiral doubling of the string 
worldsheet is taken into account. We adopt convention of the
spacetime signature to be $(-+ \cdots +)$. The $d=10$ $(32 \times 32)$ 
Dirac matrices are denoted by $\Gamma_M$. The charge conjugation $\psi_c$ 
of a spinor $\psi$ is given by
\be
\psi^{\rm T} {\cal C} \equiv \overline{\psi} = \psi^\dagger \Gamma^0.
\ee
The charge conjugation matrix $\cal C$ then satisfies the following
relations:
\be
{\cal C} \Gamma_M {\cal C}^{-1} = - {\Gamma_M}^{\rm T} ,
\qquad {\cal C}^{\rm T} = - {\cal C}.
\ee
Because the $\eta_{L,R}$ are sixteen-component Majorana-Weyl spinors, they 
satisfy both the Majorana conditions
\be
{\eta_L}^{\rm T} {\cal C} = \overline{\eta}_L, \qquad
{\eta_R}^{\rm T} {\cal C} = \overline{\eta}_R,
\ee
and the Weyl conditions
\be
\Gamma^{11} \eta_L = + \eta_L, \qquad
\Gamma^{11} \eta_R = + \eta_R.
\ee
For the present case, the $AdS_3 \times S^3 \times M^4$ background threaded
by nontrivial RR-flux is generated by a supersymmetric combination of 
ground-state $N_1$ D-strings and $N_5$ D5-branes. Suppose that 
the D1-strings are oriented along $1$-direction and the D5-branes along 
$16789$-directions, where $6789$-directions are along $M^4$. 
Due to open string sector attached to these branes, 
unbroken supersymmetry is determined by spinor parameters satisfying
\be
\Gamma^{01} \eta_L = + \eta_R , \qquad
\Gamma^{01} \eta_R = + \eta_L
\ee
from the D1-strings, and
\be
\Gamma^{01} \eta_L = + \eta_R, \qquad
\Gamma^{016789} \eta_R = + \eta_L
\ee
from the D5-branes. Combining the two conditions, we find that 
$\eta_L, \eta_R$ should satisfy 
\be
\Gamma_\perp \eta_L = + \eta_L, \qquad
\Gamma_\perp \eta_R = + \eta_R, \qquad
( \Gamma_\perp \equiv \Gamma^{6789}).
\label{condition}
\ee
Since the spinor projection is independent of transverse six-dimensional
spacetime, the chirality condition Eq.(\ref{condition}) can be rephrased 
as follows. If we decompose the sixteen-component Majorana-Weyl spinors 
$\eta_{L,R}$ into $(AdS_3 \times S_3)$ and $M_4$ parts, they are given by
\be
\eta_{L,R} = \theta^{\alpha, \alpha'} \otimes \epsilon_{L,R}^{ia} ,
\ee
where $\alpha, \alpha'$ are $Spin(3), \, Spin(3)'$ indices and $i, a$ are $
Spin(4), \, Spin''(3)$ indices. They are associated to $(AdS_3 \times S^3)$ 
and to $M_4$ times extra `isospin' spaces, respectively.
Now, for both $M^4 = T^4$ and $K_3$, residual supersymmetry of the background 
is possible only when the the spinors $\epsilon_{L,R}^{ia}$ are chirally 
projected precisely as in Eq. (\ref{condition}). 

For the case of S-dual background configuration, viz. $N_1$ F-strings
along $1$-direction and $N_5$ NS5-branes along $16789$-directions, the 
unbroken supersymmetry parameters are characterized by
\be
\Gamma^{01} \eta_L = + \eta_L, \qquad
\Gamma^{01} \eta_R = - \eta_R
\ee
due to F-string and
\be
\Gamma^{016789} \eta_L = + \eta_L,
\qquad
\Gamma^{016789} \eta_R = - \eta_R
\ee
due to NS5-brane. Combining the two conditions, one find again exactly
the same chiral projection condition as in Eq.(\ref{condition}).

In fact, the Type IIB S-duality mapping between the above two configurations
is realized on the two spinors as a $SO(2)$ rotation. To see this, it would 
be convenient to combine the two Majorana-Weyl spinors into two Weyl spinors:
\be
\eta = \eta_L + i \eta_R, \qquad \eta^* = \eta_L - i \eta_R.
\ee
In terms of the two complexified spinors, the supersymmetry projection 
condition Eq.(\ref{condition}) is rewritten as
\be
\Gamma_\perp \eta = + \eta, \quad \Gamma_\perp \eta^* = + \eta^*,
\ee
for both D1-D5 and F1-NS5 systems. Under the S-duality, the complexified 
spinors are rotated by a $U(1)$-phase:
\be
S \quad : \quad \eta \rightarrow e^{+i\pi/4} \eta,
\qquad
\eta^* \rightarrow e^{-i\pi/4} \eta^*. \label{sdual}
\ee
Actually, this is the special case of general spinor transformation rule 
under S-duality mapping \cite{ortin} :
\be
\eta \rightarrow \exp \Big( {i \over 2} {\rm arg} (c \tau + d ) \Big)
\eta
\qquad {for}
\qquad
\tau \rightarrow {a \tau + b \over c \tau + d} \quad
(ad - bc = 1) 
\label{sdual2}
\ee
and oppositely for $\eta^*$, where $\tau = {a_{\rm RR} \over 2 \pi} + 
i e^{-\phi}$ is the Type IIB dilaton. 

Closely related observation has been made in the study of the Type IIB
D-brane actions and their transformation under the $SL(2, {\bf Z})$ 
mapping on both flat \cite{aganagicetal} and $AdS^5\times S^5$\cite{duality}
spaces. Essential to the analysis of D-brane action was the fact that the 
S-duality mapping acts on the two $d=10$ Majorana-Weyl spinors as a rotation 
on $SO(2)$ subgroup of the $SL(2, {\bf R})$, the classical duality symmetry 
of Type IIB superstring.  Likewise, for similar of dual D-brane action on 
$AdS_3 \times S^3$, it should become essential to take into consideration of 
the spinor transformation rule Eqs.(\ref{sdual}, \ref{sdual2}). 

\section{ Green-Schwarz String Action on $AdS_3 \times S_3$}
We now construct the Green-Schwarz superstring action on $AdS_3 \times S^3$ via 
super-group manifold approach and utilizing $d=6$ symplectically Majorana-Weyl
spinors~\footnote{The Green-Schwarz superstring propagates on the on-shell 
background of the truncated $d=6$ $(2,0)$ supergravity theory and hence the 
string may be viewed as an ``effective'' non-critical string.}.

The $AdS_3$ space possesses $SU(1,1|2)$ invariance as a graded spacetime
symmetry, inherited  from the isometry of Type IIB D1-D5 or F1-NS5 brane
configuration. We would like to construct the Green-Schwarz superstring action 
which exhibits this graded symmetry manifestly. This is most conveniently 
achieved by viewing the the Green-Schwarz superstring action as a nonlinear 
sigma model on a super-group manifold. For the present case, the coset
superspace is ${\cal G} =
\frac{SU(1,1|2)_L \times SU(1,1|2)_R}{SO(1,2)\times SO(3)}$
whose even part is $\frac{(SU(1,1)\times SU(2))^2}{SO(1,2)\times SO(3)}
\simeq \frac{SO(2,2)}{SO(1,2)}\times \frac{SO(4)}{SO(3)}
\simeq AdS^3 \times S^3$.

The essential structure of $d=6$ supercharges are the same as in the previous
section. Starting from thirty-two component Majorana-Weyl spinors
$\eta_L, \eta_R$ in ten dimensions, one first compactify on $M_4$.
For D1-D5 brane configuration wrapped on $M_4 = T_4$ or $K_3$ (whose volume 
is taken very small), we have seen that the compactification projects each 
spinors chirally.  Low-energt dynamics on the noncompact $d=6$ spacetime is 
described precisely by (2,0) supergravity. The supergravity contains variety
of  tensor multiplets, as required by the gravitational anomaly cancellation 
(twenty-one if $M^4 = K3$). These tensor multiplets descend from NS-NS 
and R-R tensor fields in ten dimensions, but we will be restricting foregoing 
discussions only to the case for which only R-R tensor fields are turned on.

In six-dimensions with (2,0) supersymmetry, we use the  symplectically
Majorana-Weyl spinors , where the symplectic condition acts on $USp(4)=SO(5)$
R-symmetry indices $m,n$'s (and suppressing $d=6$ spinor indices):
\be
\theta^{\rm T}_m \Omega_{mn} = \overline{\theta}_n \equiv
i \theta^\dagger_n \Gamma^0.
\ee
Here, we take $d=6$ $(8\times 8)$ Dirac matrices to be antisymmetric
and the charge conjugation matrix ${\cal C}$ symmetric. As shown in the
previous section, this is achieved by decomposing the symmetric 
ten-dimensional Dirac matrices into a tensor product of antisymmetric 
$M_4$ Dirac matrices and antisymmetric $d=6$ ones. Likewise, anti-symmetric 
charge conjugation matrix can be decomposed into a tensor product of 
antisymmetric $M_4$ matrix and symmetric $d=6$ one. This way, the noncompact
six-dimensional part whose background is described by 
(2,0) supergravity becomes
completely decoupled from the internal space $M_4$.
It now remains to decompose the noncompact six-dimensional spinors further 
into direct product of spinors on $AdS_3$ and $S^3$. First, the $SO(1,5)$ 
spinor indices are decomposed into $SO(1,2) \times SO(3)$. In doing so, the 
$d=6$ R-symmetry $USp(4)$ is broken into $USp(2) \times USp(2)$, as we will 
see below.  Acting on each of the two $USp(2)$ symplectically Majorana-Weyl 
spinors, we decompose the $d=6$ Dirac matrices as
\bee
\Gamma^a &=& \gamma^a \otimes {\bf I} \otimes \sigma^1,
\qquad
{\cal C} = C_{2,1} \otimes C_3 \otimes \sigma^1 \\
\Gamma^{a'} &=& {\bf I} \otimes \gamma^{a'} \otimes \sigma^2,
\qquad
\Gamma^7 = {\bf I} \otimes {\bf I} \otimes (-\sigma^3)
\eee
Here, $C_{2,1}, C_3$ denote charge conjugation matrices for $AdS_3$
and $S^3$ spaces respectively. The resulting two-component $SO(1,2)$
spinors are Majorana and the two-component $SO(3)$ spinors are
pseudo-symplectic Majorana spinors.
In this decomposition, the symplectic-Majorana condition can be
written as
\begin{equation}
\bar{\theta}_{\alpha\alpha' \alpha''}\equiv \theta^{\beta\beta'\beta''}
C_{\alpha\beta}C'_{\alpha'\beta'}\varepsilon_{\alpha''\beta''}.
\end{equation}
The charge conjugation matrix are used in lowering and rasing the indices.
We denote $SO(1,2)$ vector indices by $a,b,c$ and $SO(3)$ vector indices
by $a',b',c'$. We use hatted indices $\hat{a}$
to denote the combination $(a, a')$.
The generators of $so(1,2)$ and $so(3)$ Clifford algebras
are $2\times 2$ matrices $\gamma_a$ and $\gamma_{a'}$
\begin{equation}
\gamma^{(a}\gamma^{b)}=\eta^{ab}=(-++), \qquad
\gamma^{(a'}\gamma^{b')}=\eta^{a'b'}=(+++).
\end{equation}
We now introduce two symplectically Majorana-Weyl supercharges $Q^I, I=1,2$.
We also denote the Pauli matrices by $\tau_i, i=1,2,3$.
Then, the $SU(1,1|2)_L\times SU(1,1|2)_R$ supersymmetry algebra
is given by
\begin{eqnarray}
\left[ P_a, P_b \right]&=&J_{ab} \nonumber \\
\left[ P_{a'}, P_{b'} \right] &=& -J_{a'b'} \nonumber \\
\left[ P_a, J_{bc} \right]&=& \eta_{ab}P_c-\eta_{ac}P_b \nonumber \\
\left[ P_{a'}, J_{b'c'} \right]&=& \eta_{a'b'}P_{c'}-\eta_{a'c'}P_{b'}
\nonumber \\
\left[J_{ab}, J_{cd}\right]&=&\eta_{bc}J_{ad}+ {\rm 3 \, terms}
\nonumber \\
\left[J_{a'b'}, J_{c'd'}\right]&=&\eta_{b'c'}J_{a'd'}+ {\rm 3 \, terms}
\nonumber  \\
\left[ Q_I, P_a\right]&=&\frac{1}{2}\tau_{3IJ}Q_J\gamma_a \label{eq:qp} \\
\left[ Q_I, P_{a'}\right]&=&\frac{i}{2}\tau_{3IJ}Q_J\gamma_{a'}
\nonumber \\
\left[ Q_I, J_{ab}\right]&=&-\frac{1}{2}Q_I\gamma_{ab} \label{eq:qj} \\
\left[ Q_I, J_{a'b'}\right]&=&-\frac{1}{2}Q_I\gamma_{a'b'}\nonumber  \\
 \{ Q_{\alpha\alpha '\alpha ''}^{I}, Q_{\beta\beta '\beta ''}^{J} \}
 &=&\delta_{IJ}(-2iC_{\alpha '\beta '}(C\gamma^a)_{\alpha\beta}P_a
+2C_{\alpha\beta}(C'\gamma^{a'})_{\alpha '\beta '}P_{a'})
\varepsilon_{\alpha ''\beta ''}  \label{eq:qq} \\
 & & -i\tau_{3IJ}(C_{\alpha '\beta '}(C\gamma^{ab})_{\alpha\beta}J_{ab}
-C_{\alpha\beta}(C'\gamma^{a'b'})_{\alpha '\beta '}J_{a'b'})
\varepsilon_{\alpha ''\beta''}.  \nonumber
\end{eqnarray}
Note that both sides of Eq.(\ref{eq:qq}) are symmetric under the exchange of
$(\alpha\alpha'\alpha'' I) \leftrightarrow (\beta\beta'\beta'' J)$
since $C$ and $C'$ are antisymmetric while $C\gamma^{a}$,
$C'\gamma^{a'}$, $C\gamma^{ab}$ and $C'\gamma^{a'b'}$ are all
symmetric. In Eq.(\ref{eq:qp}) $\tau_{3IJ}$ factor is related to that
in Eq.(\ref{eq:qq}) by the Jacobi identity.
If we define $J_a\equiv \frac{1}{2}\varepsilon_{abc}J_{bc}$ and
$J_{a'}\equiv \frac{1}{2}\varepsilon_{a'b'c'}J_{b'c'}$, then
Eq.(\ref{eq:qp}) and Eq.(\ref{eq:qj}) can be written as
\begin{eqnarray}
\left[ Q_I, J_a^J\right]&=&\frac{1}{2}\delta_{IJ}Q_J\gamma_a  \\
\left[ Q_I, J_{a'}^J\right]&=&-\frac{i}{2}\delta_{IJ}Q_J\gamma_{a'} ,
\end{eqnarray}
where $J_a^{1,2}\equiv ({J_a\pm P^a})/2$ and $J_{a'}^{2,1}\equiv 
(J_{a'}\pm P_{a'})/2$ with $P^a=\eta_{ab}P_b$.
\footnote{In proving this we have used the identities $\gamma_{ab}=
-\varepsilon_{\,\, ab}^c \gamma_c$
and $\gamma_{a'b'}=i\varepsilon_{a'b'c'}
\gamma_{c'}$.}
The resulting new variables satisfy $SU(1,1)$ and $SU(2)$ algebras
respectively:
\begin{eqnarray}
 \left[ J_{a}^I, J_{b}^J \right] &=& -\delta^{IJ}
\varepsilon_{\,\, ab}^{c}J_{c}^J
\nonumber \\
\left[ J_{a'}^I, J_{b'}^J \right] &=& -\delta^{IJ}
\varepsilon_{a'b'c'}J_{c'}^J.
\nonumber
\end{eqnarray}
Now, the Eq.(\ref{eq:qq}) can be rewritten into a transparent form: 
\begin{equation}
\{ Q_{\alpha\alpha'\alpha''}^{I}, Q_{\beta\beta'\beta''}^{J} \}
 =-4\tau_{3IJ}(iC_{\alpha'\beta'}(C\gamma^a)_{\alpha\beta}J_a^J
+C_{\alpha\beta}(C'\gamma^{a'})_{\alpha'\beta'}J_{a'}^J)
\varepsilon_{\alpha''\beta''}.
\end{equation}
viz, we have shown explicitly that the superalgebra under consideration 
is indeed $SU(1,1|2)_L \times SU(1,1|2)_R$.
From the consideration of the Jacobi identity of the form 
$\left[ Q, \{Q, Q \} \right]+$ cyclic permutations, we obtain the following
useful identity:
\begin{equation}
\gamma^a\psi_{(1}\bar{\psi}_2\gamma_a\psi_{3)}
-\gamma^{a'}\psi_{(1}\bar{\psi}_2\gamma_a^{'}\psi_{3)}=0 \label{eq:psi3}
\end{equation}
for arbitrary symplectically Majorana-Weyl spinors $\psi_i, i=1,2,3$.
This identity can be proven using the Fierz identities~\footnote{
Similar identity  can be derived from the consideration of six-dimensional
super Yang-Mills theory with the symplectically Majorana-Weyl spinors. 
The identity obtained this way has been useful in proving the 
$\kappa$-symmetry of the $d=6$ superstring action on flat background.
The same remarks should be applicable to the superstring action on 
$AdS^3 \times S^3$ background as well.} .
Moreover, in the scaling limit $P_{\hat{a}} \rightarrow R P_{\hat{a}}$,
$J_{\hat{a}}\rightarrow J_{\hat{a}}$ and $Q^I \rightarrow \sqrt{R}Q^I$
with $R\rightarrow\infty$, we obatin the supersymmetric algebra in flat
six-dimensional space. In this limit, the superstring action on 
$AdS^3 \times S^3$
should be reduced to the well-known form in the flat space.
Note that the supersymmetry algebra in flat space has $USp(4)$ R-symmetry
whereas on $AdS^3 \times S^3$ this is broken to $USp(2)\times USp(2)$.  
This is because of the term containing $\tau_{3IJ}$.

The Mauer-Cartan equations are given by
\begin{eqnarray}
dL^a&=&- L^b\wedge L^{ba}
-i\bar{L}^I\gamma^a\wedge L^I \\
dL^{ab}&=&-L^a\wedge L^b-L^{ac}\wedge L^{cb}
-i\bar{L}^I\tau_{3IJ}\gamma^{ab} L^J  \nonumber \\
 dL^{a'}&=&-L^{b'}\wedge L^{b'a'}
+\bar{L}^I\gamma^{a'}\wedge L^I \\
dL^{a'b'}&=&+ L^{a'}\wedge L^{b'}-L^{a'c'}\wedge L^{c'b'}
+i\bar{L}^I\tau_{3IJ}\gamma^{a'b'} L^J \nonumber \\
dL^I&=& + \frac{1}{2} \tau_3^{IJ} \gamma^aL^J\wedge L^a
+\frac{1}{4}\gamma^{ab}L^I\wedge L^{ab}
+\frac{i}{2} \tau_3^{IJ} \gamma^{a'}L^J\wedge L^{a'}
+\frac{1}{4}\gamma^{a'b'}L^I\wedge L^{a'b'} \\
d\bar{L}^I&=&-\frac{1}{2} \tau_3^{IJ} \bar{L}^J\gamma^a\wedge L^a
-\frac{1}{4}\gamma^{ab}\bar{L}^I\wedge L^{ab}
-\frac{i}{2} \tau_3^{IJ} \bar{L}^J\gamma^{a'}\wedge L^{a'}
-\frac{1}{4}\bar{L}^I\gamma^{a'b'}\wedge L^{a'b'}. \nonumber
\end{eqnarray}

Consider the expression
${\cal H}_1\equiv A^{IJ}L^a\bar{L}^I\gamma^a L^J$. We find that
\begin{eqnarray}
d{\cal H}_1&=&-i\bar{L}^K\gamma^a L^K A^{IJ}\bar{L}^I \gamma_a L^J
+\frac{1}{2}(\tau_3 A+A\tau_3)_{IK}L^a L^b\bar{L}^{I}\gamma^{ab}L^K \\
  & & +\frac{i}{2}(A\tau_3-\tau_3 A)^{IK}L^a L^{a'}
\bar{L}^{I}\gamma^a\gamma^{a'}L^K. \nonumber
\end{eqnarray}
Similarly, for ${\cal H}_2\equiv A^{'IJ}L^{a'}\bar{L}^I\gamma^{a'} L^J$,
we have
\begin{eqnarray}
d{\cal H}_2&=&\bar{L}^K\gamma^{a'} L^K A^{'IJ}\bar{L}^I \gamma_{a'} L^J
+\frac{1}{2}(\tau_3 A'+A'\tau_3)_{IK}L^{a'} L^{b'}\bar{L}^{I}
\gamma^{a'b'}L^K \\
  & & +\frac{1}{2}(A'\tau_3-\tau_3 A')^{IK}L^{a'} L^a
\bar{L}^{I}\gamma^a\gamma^{a'}L^K. \nonumber
\end{eqnarray}
If we choose $A, A'$ to be anticommuting with $\tau_3$ and $A'=iA$,
we find that all terms cancel except for the four-fermion terms.
Moreover, if $A$ is proportional to $\tau_2$, then ${\cal H}_1$ and ${\cal H}_2$
themselves vanish as $\bar{L}^I\gamma^{\hat{a}}L^J
=\bar{L}^J\gamma^{\hat{a}}L^I$.
Thus, the only choice is that $A$ is proportional to $\tau_1$.
For this choice, the four-fermion terms also vanish due to the identity
Eq.(\ref{eq:psi3}).

Thus the Green-Schwarz action on $AdS_3 \times S^3$ is given by
\be
I=-\frac{1}{2}\int _{\partial M_3}d^2\sigma \sqrt{-g}g^{ij}
(L^a_i L^a_j +L^{a'}_i L^{a'}_j)+i\int_{M_3} {\cal H} ,
\ee
where $\cal{H}$ is a closed 3-form invariant under
 $SO(1,2)\times SO(3)$:
\be
{\cal H}=\tau_1^{IJ}(L^a\wedge\bar{L}^I \gamma^a\wedge L^J
+i L^{a'}\wedge\bar{L}^I \gamma^{a'}\wedge L^J).
\ee
The relative coefficient between the usual kinetic term and the
Wess-Zumino term is fixed by the $\kappa$-symmetry.
Under the variation $\delta\theta^I$,
\begin{eqnarray}
\delta L^a&=&2i\bar{L}^I \gamma^a \delta\theta^I \\
\delta L^{a'}&=&-2\bar{L}^I \gamma^{a'} \delta\theta^I \nonumber \\
\delta L^I&=&d\delta \theta^I+\frac{1}{2}\tau_{3IJ}
(L^a\gamma^a\delta\theta^J+iL^{a'}\gamma^{a'}\delta\theta^J) \nonumber \\
 & &+\frac{1}{4}L_{ab}\gamma^{ab}\delta\theta^I
+\frac{1}{4}L^{'}_{ab}\gamma^{ab'}\delta\theta^I. \nonumber
\end{eqnarray}
The variation of ${\cal H}$ is given by $\delta{\cal H}=d\Lambda$, where
\begin{equation}
\Lambda=\tau_{1IJ} \, \left(2L^a\bar{L}^I\gamma^a\delta\theta^I
+2iL^{a'}\bar{L}^I\gamma^{a'}\delta\theta^I \right).
\end{equation}
In order to identify the $\kappa$-symmetry, we have found it is useful to 
define ``rotated'' variables :
\begin{equation}
\theta^{1'}\equiv \frac{\theta^1+\theta^2}{\sqrt{2}} \qquad \quad
\theta^{2'}\equiv \frac{-\theta^1+\theta^2}{\sqrt{2}}
\end{equation}
and analogous expressions for other spinors.

If we define $\delta x^a\equiv \delta X^M L^a_M$, 
$\delta x^{a'}\equiv \delta X^M L^{a'}_M$, and $ \delta\theta^I\equiv
\delta X^M L^I_M$, the $\kappa$-symmetry transformation is given by
\begin{eqnarray}
\delta_{\kappa}x^a \,\,\, &=& \,\,\, 0 
 \\
\delta_{\kappa}x^{a'} \,\,\, &= & \,\,\, 0,
\\
\delta_{\kappa}\theta^{I'}&=&2 \,  
\left( L_i^a\gamma^a-iL_i^{a'} \gamma^{a'} \right) \, \kappa^{iI'} \\
\delta_{\kappa}(\sqrt{g}g^{ij})&=&-16i\sqrt{g}
\left( \, P_{-}^{jk}\bar{L}_k^{1'}\kappa^{i1'}+
P_{+}^{jk}\bar{L}_k^{2'}\kappa^{i2'} \right)  \nonumber ,
\end{eqnarray}
where $P_{\pm}=\frac{1}{2}(g^{ij}\pm \frac{1}{\sqrt{g}}\varepsilon^{ij})$.
We see that the eigenbasis for the $\kappa$-symmetry does not coincide
with the diagonal basis for the superconformal algebra.

The gauge-fixing of the $\kappa$-symmetry may be proceeded similarly as in
Ref.~\cite{kallosh}. The algebraic steps and results
are essentially the same, and will not be repeated here.
However, it's not clear if this Killing spinor gauge fixing provides
any useful insight to understand the quantization of the theory.

Using the above super-coset approach, we can find the closed form
expression for the supervielbein by solving the Mauer-Cartan equations
Eqs.(27 - 29). For $AdS^5\times S^5$, such an expression has been found 
in \cite{kallosh4}.
Similar analysis was made for M2-brane on $AdS^4\times S^7$ or 
$AdS^7 \times S^4$ \cite{plefka}. In \cite{plefka}, the result obtained from 
the super-coset approach was also compared to the construction of the 
super-space vielbein in terms of the component field of eleven-dimensional 
on-shell supergravity. The two results turn out to agree perfectly. 
For the $d=2$ (2,0) supergravity, we expect essentially the same result.
The only difference is that the $d=6$ (2,0) supergravity is described in 
$SO(5, N_T)$ invariant way for the tensor multiplet coupling \cite{romans}.
 For example, if we consider the Type IIB compactification on $K3$, 
the resulting theory has $SO(5,21)$ invariance which rotates the tensor 
multiplets coming from the compactification.
If we compare the Killing spinor equation Eq.(16) in \cite{sundell}
with our expression on the commutator [Q, P] in Eq.(20), we see that in our
expression the $SO(5, N_T)$ symmetry is broken by choosing a particular
direction in the $SO(5, N_T)$ space. 

So far, we have considered the $AdS^3 \times S^3$ with R-R 3-form field. 
In general, the background 3-form flux can be a mixture of 
NS-NS and R-R field.  If the $AdS^3 \times S^3$ background includes
the NS-NS 3-form flux, in the Wess-Zumino term, we  should 
have an additional term proportional to $L^a \wedge L^b \wedge L^c H_{abc}$.
If the 3-form flux comes entirely from NS-NS field, the corresponding 
action with this additional term is nothing but the Green-Schwarz formulation
of the Type IIB string considered by \cite{giveonkutasovseiberg} using
the Neveu-Schwarz-Ramond formalism. It would be very intersting if one can
establish equivalence between the Green-Schwarz formalism and the 
Neveu-Schwarz-Ramond formalism for $AdS^3 \times S^3$ background with pure 
NS-NS 3-form flux.

\section{Conclusion}
In this paper, we have constructed the Green-Schwarz superstring action on
$AdS_3 \times S^3$, where the $S^3$ is threaded by pure Ramond-Ramond tensor
flux. Essential to the construction was to utilize the symplectically 
Majorana-Weyl representation of the $d=6$ supercharges. 
The representation makes it most transparent and straightforward to compare
the $\kappa$-symmetry constraint to on-shell background of the standard 
$(2,0)$ supergravity. 



\begin{thebibliography}{1}

\bibitem{malda} J. Maldacena, Adv. Theor. Math. Phys. {\bf 2} (198) 231.

\bibitem{MeTs} R.R. Metasev and A.A. Tseytlin, {\sl Type IIB
Superstring Action in $AdS_5 \times S^5$ background}, {\tt hep-th/9805028};
\\
I. Pesando, {\sl All Roads Lead to Rome:
Supersolvables and Supercosets}, {\tt hep-th/9808146};\\
I. Pesando, {\sl A $\kappa$ Gauge Fixed Type IIB
Superstring Action on $AdS_5 \times S^5$}, {\tt hep-th/9808020};\\
R. Kallosh and A.A. Tseytlin, {\sl
Simplifying Superstring Action on $AdS_5 \times S^5$}, {\tt hep-th/9808088}.

\bibitem{kallosh} R. Kallosh, {\sl Superconformal Actions on Killing Gauge},
{\tt hep-th/9807206};\\
R. Kallosh and J. Rahmfeld, {\sl The GS Action on $AdS_5 \times S^5$}, 
{\tt hep-th/9808038}.

\bibitem{adsdbrane}
P. Claus, {\sl Super M-Brane Actions in $AdS_4 \times S^7$ and
$AdS_7 \times S^4$}, {\tt hep-th/9809045};\\
I. Oda, {\sl Super D-String Action on $AdS_5 \times S^5$}, {\tt
hep-th/9809076};\\
P. Pasti, D. Sorokin, M. Tonin, {\sl On Gauge-Fixed
Superbrane Actions in AdS Super-backgrounds}, {\tt hep-th/9809213}.

\bibitem{plefka} B. de Wit, K. Peeters, J. Plefka and A. Sevrin,
{\sl The M-theory Two-Brane in $AdS^4 \times S^7$ and
$AdS^7 \times S^4$}, {\tt hep-th/9808052} .

\bibitem{duality}
I. Oda, {\sl $SL(2,Z)$ Self-duality of Super D3-Brane
Action on $AdS_5 \times S^5$}, {\tt hep-th/9810024};\\
T. Kimura, {\sl Self-Duality of Super D3-brane
Action on $AdS_5 \times S^5$ Background}, {\tt hep-th/9810136};\\
J. Park and S.-J. Rey, {\sl Dual D-Brane Actions on $AdS_5 \times S^5$},
{\tt hep-th/9810154}.

\bibitem{giveonkutasovseiberg} A. Giveon, D. Kutasov and N. Seiberg,
{\sl Comments on String Theory on $AdS_3$}, {\tt hep-th/980619}.

\bibitem{pesando} I. Pesando, {\sl The GS Type IIB Superstring Action
on $AdS_3 \times S^3 \times T^4$}, {\tt hep-th/9809145};\\
J. Rahmfeld and A. Rajaraman, {\sl
The GS String Action on $AdS_3 \times S^3$ with Ramond-Ramond Charge},
{\tt hep-th/9809164} .

\bibitem{romans} L. Romans, Nucl. Phys. {\bf B276} (1986) 71.

\bibitem{boonstra} H. Boonstra, B. Peeters and K. Skenderis, Nucl. Phys
{\bf B533} (1998) 127.

\bibitem{kallosh3} P. Claus, R. Kallosh, J. Kumar, P. Townsend and
A. Van Proeyen, J. High Energy Phys. {\bf 06} (1998) 4.

\bibitem{ortin} T. Ortin, Phys. Rev. {\bf D51} (1995) 790, {\tt hep-th/
9404035}.


\bibitem{aganagicetal} M. Aganagic, J. Park, C. Popescu and J.H.
Schwarz, Nucl. Phys. {\bf B496} (1997) 215.


\bibitem{kallosh4} R. Kallosh, J. Rahmfeld and A. Rajaraman, {\sl Near
Horizon Superspace}, {\tt hep-th/9805217}.

\bibitem{sundell}S. Deger, A. Kaya, E. Sezgin and P. Sundell,
{\sl Spectrum of D=6, N=4b Supergravity on $AdS^3 \times S^3$},
{\tt hep-th/9804166}.

\end{thebibliography}
\end{document}